# Observation of surface solitons in chirped waveguide arrays

A. Szameit[1], Y. V. Kartashov[2], F. Dreisow[1], M. Heinrich[1], T. Pertsch[1], S. Nolte[1], A. Tünnermann[1], V. A. Vysloukh[2], and L. Torner[2]

[1]Institute of Applied Physics, Friedrich Schiller University Jena, Max-Wien-Platz 1, 07743 Jena, Germany

[2]ICFO-Institut de Ciencies Fotoniques, and Universitat Politecnica de Catalunya, Mediterranean Technology Park, 08860 Castelldefels (Barcelona), Spain

We report the observation of surface solitons in chirped semi-infinite waveguide arrays whose waveguides exhibit exponentially decreasing refractive indices. We show that the power threshold for surface wave formation decreases with an increase of the array chirp and that for sufficiently large chirp values linear surface modes are supported.

Surface states at the interfaces of uniform and periodic nonlinear materials have attracted considerable attention because of their unique physical properties [1]. One-dimensional surface solitons in focusing lattices were predicted recently [2,3]. Surface lattice solitons in defocusing [4-7] and in quadratic [8] media were also analyzed. Very recently surface waves were observed at interfaces of two-dimensional optically induced [9] and laser-written [10] lattices. Surface solitons may also form at interfaces of materials with more complex periodic refractive index landscapes [11,12]. Deviations from a strict periodicity may lead to new interesting phenomena. While chirping of infinite waveguide array results in unique dynamics [13], in the presence of an interface, it may lead to appearance of thresholdless surface states [14,15]. In this Letter we report on the experimental observation of surface waves in fs-laser written waveguide arrays with an exponentially chirped refractive index. We reveal a clear dependence of the power threshold for surface wave formation on the array chirp.

Our theoretical model is based on the nonlinear Schrödinger equation for the spatial dynamics of the dimensionless field amplitude $q$:

$$i\frac{\partial q}{\partial \xi} = -\frac{1}{2}\frac{\partial^2 q}{\partial \eta^2} - |q|^2 q - pR(\eta)q. \qquad (1)$$

Here the transverse coordinate $\eta$ and the longitudinal coordinate $\xi$ are normalized to the beam width and diffraction length; the parameter $p$ describes the refractive index modulation depth and the function $R(\eta) = \sum_{m=0}^{\infty} \exp[-(\eta - mw_s)^8 / w_\eta^8]\exp(-\alpha m)$ stands for the refractive index profile, where $w_s$ is the waveguide separation, $w_\eta$ is the waveguide width, and $\alpha$ is a measure for the array chirp. In the simulations the beams $A\exp(-\eta^2/W^2)$ with width $W = 0.3$ were used for the excitation centered at the border waveguide.

The experiments were accomplished in fs laser written waveguide arrays, whose fabrication parameters can be found elsewhere [16]. Since the refractive index change of the individual guides crucially depends of the writing velocity [17], it is possible to introduce a chirp in the waveguides by changing their writing speed. We fabricated two classes of waveguide arrays, one having a separation of 13 $\mu$m ($w_s = 1.3$) and the other 40 $\mu$m ($w_s = 4$). In all our samples the width of individual guides was 3 $\mu$m ($w_\eta = 0.3$). The nonlinearity is focusing ($n_2 = 2.7 \times 10^{-20}$ m$^2$/W).

To illustrate the impact of the array chirp on the propagation of low-power excitations we selected the array with 13 $\mu$m separation and launched light at $\lambda = 633$ nm into the surface waveguide. The refractive index change in this waveguide is $\sim 7.6 \times 10^{-4}$ that corresponds to $p = 11$ in Eq. (1). The evolution inside the array was directly monitored by detecting the fluorescence of the propagating light [18]. In the unchirped case discrete diffraction results in recession of light from the interface into the array depth [Fig. 1(a)]. The refractive index of the boundary waveguide slightly differs from other waveguides due to writing procedure, where the material in the vicinity of the waveguide is affected too. Hence, one obtains a small detuning of the boundary guide which has only one neighbor instead of two, causing the appearance of a near-surface defect that can capture a small portion of the input energy. This effect however is almost negligible in comparison with the chirping. Even in the presence of a small chirp ($\alpha = 0.014$) the distributed reflection from the array causes partial regression of radiation towards the interface, so that one can observe near surface light oscillations [Fig. 1(b)]. The frequency of these oscillations gradually increases with increase of $\alpha$, while the degree of penetration of radiation into the array decreases. Particular representations of this behavior are de-picted in Figs. 1(c) and 1(d) for chirp values of $\alpha = 0.028$ and $\alpha = 0.042$, respectively. This indicates the formation of linear surface modes for sufficiently high chirps.

To understand the impact of chirping on soliton properties we selected another array with $40\,\mu\text{m}$ that is suitable for the investigation of high-power excitations. This array features a refractive index modulation $\sim 2.6 \times 10^{-4}$ that corresponds to $p = 2.3$ at $\lambda = 800\,\text{nm}$. The increased waveguide spacing compared to the sample for the linear experiments is mainly caused by the limitation of the applied peak powers by the damage threshold of the material. In order to generate a nonlinear surface state, we had to reduce the required peak power below this threshold, which was achieved by reducing the linear coupling strength between the individual guides. An appropriate waveguide spacing is $40\,\mu\text{m}$, which was used in previous experiments [16].

First we study theoretically stationary solitons in Eq. (1) of the form $q(\eta,\xi) = w(\eta)\exp(ib\xi)$, where $b$ is the propagation constant. The properties of the derived solitons are summarized in Fig. 2. At high energy flows $U = \int_{-\infty}^{\infty} |q|^2 \, d\eta$ (or $b$ values) solitons are well localized in both unchirped [Fig. 2(a)] and chirped [Fig. 2(b)] arrays. The soliton shapes are modulated: The local intensity maxima coincide with the waveguide centers. Decreasing $b$ causes a gradual increase of the soliton width and an expansion into array depth that is considerable in the unchirped array [Fig. 2(a)], but less pronounced in chirped array [Fig. 2(b)], especially for large $\alpha$. This difference in soliton shapes finds its manifestation in the qualitatively different behavior of the energy flows close to the cutoff $b_{\text{co}}$. Thus, in the unchirped array $U$ abruptly diverges as $b \to b_{\text{co}}$ so that surface solitons exist only above the threshold $U_{\text{th}}$, while in a chirped array $U$ may vanish at the cutoff provided that $\alpha$ is sufficiently large [Fig. 2(e)]. Therefore, under appropriate conditions chirped arrays support thresholdless surface waves which exist for $b \geq b_{\text{co}}$, where $b_{\text{co}}$ coincides with the propagation eigenvalue of linear guided mode. The properties of such linear modes are described in Fig. 2(c) showing the number of waveguide where the mode maximum is located as a function of the chirp. One can see that increasing the chirp causes a displacement of the center of the linear mode from the array depth towards the surface. When $\alpha > \alpha_{\text{cr}}$ (at $p = 2.3$ one has $\alpha_{\text{cr}} \approx 0.0095$) a linear surface mode exists and all surface solitons residing in the first channel do not require a threshold energy flow for their existence (note, that an increasing $p$ causes a monotonic de-crease of the critical chirp value [Fig. 2(d)]). In contrast, when $\alpha < \alpha_{\text{cr}}$ linear guided modes do not exist and surface solitons form only above a threshold energy flow. This threshold energy flow is a monotonically decreasing function of $\alpha$ [Fig. 2(f)], which implies that an increasing chirp should also facilitate a dynamical excitation of surface modes.

These findings are confirmed by our experiments. In Fig. 3, we compare the experimental results with numerical simulations using Eq. (1) in the ideal case of CW illumination. Since the experiments were conducted with pulsed light ($\tau_{\text{pulse}} = 150\,\text{fs}$), one can make only a qualitative comparison between the experiments and simulations to confirm the consistency of the observations with the fact of soliton formation around the pulse peak. In Fig. 3, every subplot consists of a theoretical part, showing the propagation dynamics inside the sample, on top of the photograph of the experimentally observed output patterns. For an unchirped array ($\alpha = 0$), at $50\,\text{kW}$ input peak power the light penetrates into the array [Fig. 3(a), top row]. This corresponds to the linear case, where no bounded surface mode exists. For an increased input power of $500\,\text{kW}$, the light starts to localize in the excited surface waveguide, which is shown in Fig. 3(a), second row. At $700\,\text{kW}$, the light is even more localized in the surface waveguide [Fig. 3(a), third row]. Finally, at $1200\,\text{kW}$ input power, a discrete surface soliton has been excited. This is supported by simulations using the input condition $A\exp(-\eta^2/W^2)$ with the amplitudes $A = 0.23$, $0.74$, $0.86$, and $1.22$ respectively. The dynamics of the excitation in such an unchirped array suggests that a non-vanishing threshold $U_{\text{th}} > 0$ exists, so that for $U < U_{\text{th}}$ no stable surface mode exists. The picture completely changes, when the array exhibits a chirp of $\alpha = 0.048$. In this case already a linear localized surface mode exists for a low input power of $50\,\text{kW}$ [Fig. 3(b), top row]. When the power is increased [Fig. 3(b), second row [$500\,\text{kW}$], third row [$700\,\text{kW}$], fourth row [$1200\,\text{kW}$]), the localization is further increased. However, since already in the low power limit a stable surface mode can be observed, the threshold for the soliton formation vanishes ($U_{\text{th}} = 0$). This is fully consistent with our theory.

In conclusion, we experimentally demonstrated the existence of stable linear surface modes in chirped fs laser written waveguide arrays. We found a vanishing energy threshold for the formation of discrete surface solitons in lattices with a sufficiently high chirp, which was also confirmed by our experiments.

# References with titles

# References without titles

**Figure captions**

Figure 1.   Comparison of experimental (top row) and theoretical (bottom row) intensity distributions for low-power excitation of the first waveguide of the array. The upper edge of each panel corresponds to the input sample facet; while the lower edge corresponds to the output facet. Panels (a) correspond to unchirped array, in (b) $\alpha = 0.014$, in (c) $\alpha = 0.028$, and in (d) $\alpha = 0.042$.

Figure 2.   (a) Profiles of solitons corresponding to $b = 0.671$ (curve 1, black) and $b = 0.617$ (curve 2, red) at $\alpha = 0$, $p = 2.3$. (b) Profiles of solitons corresponding to $b = 0.671$ (curve 1) and $b = 0.594$ (curve 2) at $\alpha = 0.02$, $p = 2.3$. The gray waveguide regions are defined by $R(\eta) > 1/2$. (c) The number of waveguide where linear mode maximum is located versus $\alpha$ at $p = 2.3$. (d) Critical chirp for linear surface mode existence versus $p$. (e) Energy flow versus $b$ at $p = 2.3$ for $\alpha = 0$ (curve 1) and $\alpha = 0.02$ (curve 2). Points marked by circles correspond to the profiles shown in (a) and (b). (f) Soliton threshold energy flow versus $\alpha$ at $p = 2.3$.

Figure 3.   Comparison of experimental and theoretical intensity distributions for the excitation of the first waveguide of the array. Gray-scale theoretical plots, showing propagation dynamics inside the sample, are placed on top of photographs showing experimental intensity distributions at the output sample facet. Figure (a) corresponds to an unchirped array, in (b) the chirp is $\alpha = 0.048$. In all cases the first row corresponds to a peak power of $50\,\text{kW}$, the second row corresponds to $500\,\text{kW}$, the third row to $700\,\text{kW}$, and last row to $1200\,\text{kW}$. In simulations the amplitudes of the input Gaussian beams were $A = 0.23$, $0.74$, $0.86$, and $1.22$, respectively.

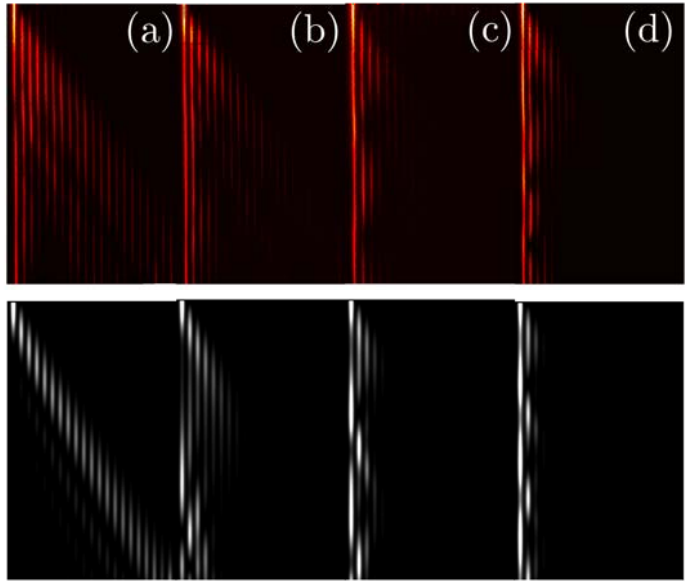

Figure 1. Comparison of experimental (top row) and theoretical (bottom row) intensity distributions for low-power excitation of the first waveguide of the array. The upper edge of each panel corresponds to the input sample facet; while the lower edge corresponds to the output facet. Panels (a) correspond to unchirped array, in (b) $\alpha = 0.014$, in (c) $\alpha = 0.028$, and in (d) $\alpha = 0.042$.

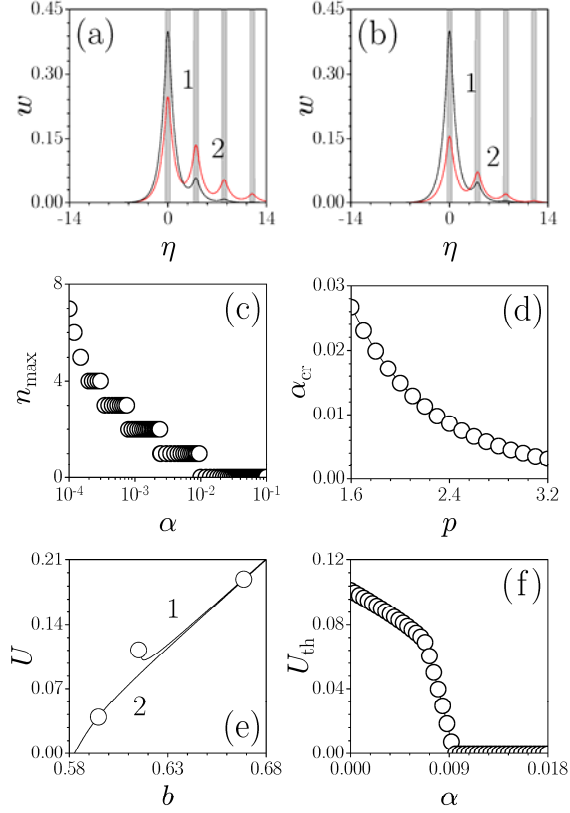

Figure 2. (a) Profiles of solitons corresponding to $b = 0.671$ (curve 1, black) and $b = 0.617$ (curve 2, red) at $\alpha = 0$, $p = 2.3$. (b) Profiles of solitons corresponding to $b = 0.671$ (curve 1) and $b = 0.594$ (curve 2) at $\alpha = 0.02$, $p = 2.3$. The gray waveguide regions are defined by $R(\eta) > 1/2$. (c) The number of waveguide where linear mode maximum is located versus $\alpha$ at $p = 2.3$. (d) Critical chirp for linear surface mode existence (in border guide) versus $p$. (e) Energy flow versus $b$ at $p = 2.3$ for $\alpha = 0$ (curve 1) and $\alpha = 0.02$ (curve 2). Points marked by circles correspond to the profiles shown in (a) and (b). (f) Soliton threshold energy flow versus $\alpha$ at $p = 2.3$.

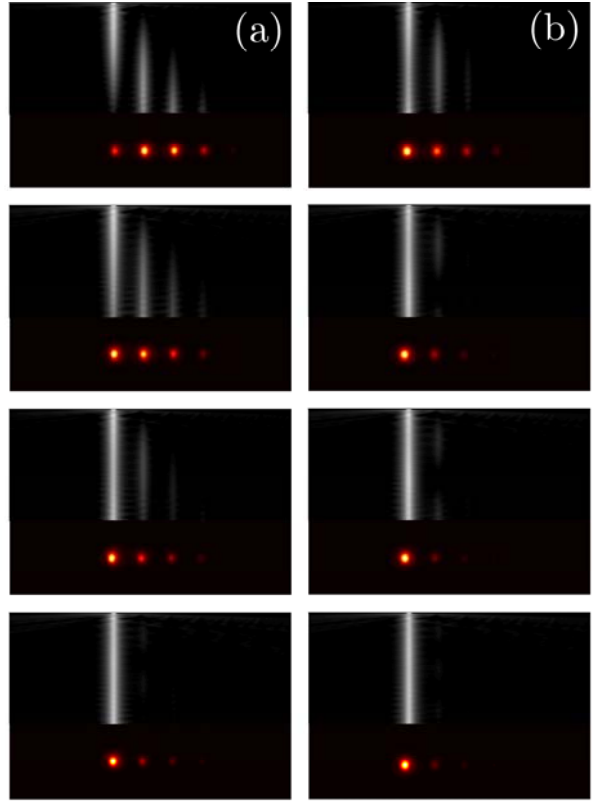

Figure 3.   Comparison of experimental and theoretical intensity distributions for the excitation of the first waveguide of the array. Gray-scale theoretical plots, showing propagation dynamics inside the sample, are placed on top of photographs showing experimental intensity distributions at the output sample facet. Figure (a) corresponds to an unchirped array, in (b) the chirp is $\alpha = 0.048$. In all cases the first row corresponds to a peak power of 50 kW, the second row corresponds to 500 kW, the third row to 700 kW, and last row to 1200 kW. In simulations the amplitudes of the input Gaussian beams were $A = 0.23$, 0.74, 0.86, and 1.22, respectively.